\documentclass[preprint2]{aastex}

\newcommand{\cmd}{CMD}

\shorttitle{NGC 6866}
\shortauthors{Janes, et al.}

\begin{document}

\title{Open Clusters in the Kepler Field, II. NGC 6866}

\author{Kenneth Janes}
\affil{Astronomy Department, Boston University, Boston, MA 02215}
\author{Sydney A. Barnes$^*$}
\affil{Leibniz-Institute for Astrophysics, Potsdam, Germany}
\altaffiltext{*}{Also Space Science Institute, Boulder, CO 80301}
\author{S\o ren Meibom}
\affil{Harvard-Smithsonian Center for Astrophysics, 60 Garden St., 
Cambridge, MA 02138}
\and
\author{Sadia Hoq}
\affil{Astronomy Department, Boston University, Boston, MA 02215}

\begin{abstract}

We have developed a maximum-likelihood procedure to fit theoretical 
isochrones to the observed cluster color-magnitude diagrams of NGC 6866, an
open cluster in the Kepler Spacecraft field of view.
The Markov-Chain
Monte Carlo algorithm permits exploration of the entire parameter space of 
a set of isochrones to find both the best solution and the 
statistical uncertainties. For clusters in the age range of NGC 6866, with 
few if any red giant members, a purely 
photometric determination of the cluster properties is not well-constrained.
Nevertheless, based on our UBVRI photometry alone, we have derived 
the distance, reddening, age and metallicity of the cluster 
and established estimates for the binary nature and membership probability of 
individual stars.  We derive the following values for the cluster properties:
$(m-M)_V = 10.98\pm 0.24$, $E(B-V) = 0.16\pm 0.04$ (so the distance = 1250 pc),
age $= 705\pm170$\ Myr and
$Z = 0.014\pm 0.005$.

\end{abstract}


\section{Introduction}

The four open clusters located in NASA's Kepler spacecraft field of 
view (NGC 6866, NGC 6811, NGC 6819 and NGC 6791), which range in age from 
less than 1 Gyr to more than 8 Gyr,  are ideal targets for a variety of 
astrophysical investigations. The two older clusters, NGC 6819 and NGC 6791
have been thoroughly studied \citep[see, e.g.,][for recent Kepler-related
papers]{corsaro12, sandquist13, 
yang13}, but the other two are less well-known.
We recently completed an analysis of NGC 6811 
\citep[][hereafter Paper I]{janes13} and we report here on NGC 6866. 

Recent publications on 
NGC 6866 include \citet{frolov10}, who did a proper motion 
and CCD photometric study of the cluster; photometric studies by
\citet{molenda09} and \citet{joshi12}; and an investigation by
\citet{balona13} of rotations and
pulsations of stars in the cluster using Kepler data.  
Frolov et al. found 423 likely cluster 
members and estimated the age of the cluster at 560 Myr.
Joshi et al. derived $E(B-V) = 0.10$\ mag., an age of 630 Myr and a distance
of 1.47 kpc.  From an analysis of 2MASS JHK photometry, \citet{gunes12}
derived an age 0.8$\pm$0.1 Gyr, E(B-V) = 0.19$\pm$0.06 and $(m-M)_{\circ} = 
11.08\pm0.11$.
Molenda-\.Zakowicz et al. and Joshi et al. both found 
several variable stars in the cluster.  In a survey of a number
of clusters, \citet{frinchaboy08} derived a radial velocity of 12.18 $\pm$\ 1.14
km s$^{-1}$ and proper motions $\mu_{\alpha}cos\delta = -5.52 \pm 1.17$\ 
mas yr$^{-1}$ and $\mu_{\delta} = -7.97 \pm 1.09$\ mas yr$^{-1}$\ for the
cluster.

\begin{figure}
\epsscale{1.05}
\plotone{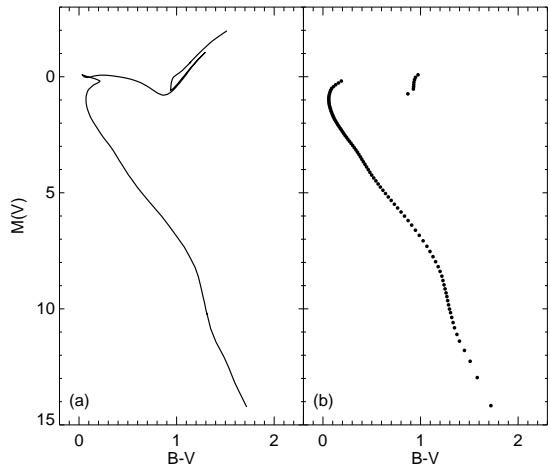}
\caption{(a) A typical isochrone \citep{bressan12} with age = 700 Myr and
Z = 0.014 (see Table \ref{final}),  (b)  
A theoretical 
CMD consisting of 100 stars selected at regular mass intervals from the same 
isochrone. We use the theoretical CMD for our analysis rather than the 
isochrone.
\label{theory}}
\end{figure}

The traditional approach to deriving the properties of star clusters from 
broad-band photometry is to ``fit'' theoretical stellar isochrones to the
observed color-magnitude diagram (CMD).  But an isochrone is simply the locus 
of possible stars over a range of masses at fixed age in the CMD (see 
Figure \ref{theory}a).  
So at a given time, in a cluster with a finite number of stars, 
it is unlikely that any stars will actually be found along some sections of an
isochrone. 

We have chosen instead to find the maximum-likelihood probability that the 
observed cluster stars embedded in a background population of field stars 
(Figure \ref{allbv}a) 
could be drawn from a theoretical CMD, created by choosing a 
sample of stars from the models with a range of masses, as in Figure
\ref{theory}b.  

By posing the problem in this way, we can take a modern Bayesian approach to 
derive not only the cluster parameters, but also their 
uncertainties.  

In \S 2 we describe the photometry; \S 3 is a discussion of the structural
properties of the cluster; The Bayesian analysis is the topic of \S 4;
\S 5 is a discussion of our results; and \S 6 summarizes our conclusions.

\section{Photometric Program}

We acquired CCD images of the cluster and standard stars
on four nights (see Table \ref{log}), with 
the 1.2-meter Hall and 1.8-meter Perkins telescopes at Lowell Observatory. 
Since we did all of our NGC 6866 observing
on nights when we also observed NGC 6811, the details of the observing and
photometric reductions are given in Paper I. The following is a summary of the 
observational program.

\begin{deluxetable}{lcccc}
\tabletypesize{\footnotesize}
\tablewidth{0pt}
\tablecaption{Observing Record\label{log}}
\tablehead{\colhead{Date} & \colhead{2009 Sept 21} & \colhead{2009 Sept 25} &
\colhead{2010 Sept 17}  & \colhead{2010 Sept 18}
}
\startdata

Telescope & Perkins 1.8m & Hall 1.2m & Hall 1.2m & Hall 1.2m \\
Detector  & Fairchild CCD & SITe CCD & e2v CCD & e2v CCD \\
Filters   & B,V,R,I       & B,V,I    & U,B,V   & U,B,V \\
No. of Stds. & 93, 101, 96, 96 & 42, 34, 33 & 150, 129, 137 & 181, 188, 208 \\
Field Size & 13.3$^{\prime}$ $\times$ 13.3$^{\prime}$ &  
37.9$^{\prime}$ $\times$ 37.9$^{\prime}$ & 
42.8$^{\prime}$ $\times$ 42.8$^{\prime}$ & 
42.8$^{\prime}$ $\times$ 42.8$^{\prime}$ \\

\enddata
\end{deluxetable}

Conditions were photometric on all four 
nights, and we imaged extensive sequences of Landolt standard stars 
\citep{landolt09} over a 
wide range of airmass on each night, as noted in Table \ref{log}.  
We used IRAF\footnote{IRAF is distributed by NOAO, operated by AURA 
under a 
cooperative agreement with the NSF.}   
bias subtraction and flat fielding functions for our initial 
image processing.  
We used the SPS program \citep{janes93} 
for PSF-fitting photometry and we transformed the photometry to 
the UBVRI colors as described in Paper I.
The calculated uncertainties in the transformation coefficients are all less 
than 0.01 mag.  The standard deviations of the nightly B and V values from 
the four-night means and the average differences between 
the two nights each in U 
and I photometry are all within 0.01 magnitude.

To aid in merging the measurements from individual frames, we used stars from 
the Kepler Input Catalog \citep{brown11} to develop
preliminary
transformation coefficients to 
the equatorial system relative to the assumed cluster center at  RA = 
20$^h$\ 03$^m$\ 55.0$^s$, Dec = +44$^{\circ}$\ 09$^{\prime}$\ 
30.0$^{\prime\prime}$.  
We assumed that star 
images with transformed 
coordinates within 0.5 arcseconds of one another on different frames
are the same star, except that we 
rejected stars if a single star image on one frame was within 0.5 
arcseconds of two stars on another frame.  We also rejected stars with
fewer than 3 B or V measurents or fewer than 2 R or I measurements or if
the calculated standard deviation of the V magnitude or B-V color index is 
0.1 magnitude or greater.  

The resulting catalog (Table \ref{catalog}) contains 7714 stars.  Star numbers
are in column 1, positions in arcseconds relative to the center 
($X = \alpha cos(\delta)$\ and $Y = \delta$) are
in columns 2 and 3, the $V$\ magnitudes and errors and $B - V$, $U - B$, 
$V - R$\ and $V - I$\ color indices and their errors are in the next 10 
columns, and the final columns contain the numbers of measurements in each
filter (in the order U, B, V, R, I).  For the final catalog, we tied the
preliminary merged star positions aproximately to
the UCAC4 coordinate system \citep{zacharias13}.

\begin{deluxetable}{rrrcccccrrrrr}
\tabletypesize{\small}
\tablewidth{0pt}
\tablecaption{Photometric Catalog \label{catalog}}
\tablehead{\colhead{Star}& \colhead{X }&\colhead{Y }&\colhead{V }
&\colhead{B-V}&\colhead{U-B}&\colhead{V-R}&\colhead{V-I}
&\colhead{N$_U$}&\colhead{N$_B$}&
\colhead{N$_V$}&\colhead{N$_R$}&\colhead{N$_I$}
}

\startdata

1 & -1267.53 & 1151.81 & 14.642 $\pm$ 0.006 & 1.846 $\pm$ 0.014 & \nodata 
& \nodata & \nodata & 0 & 3 & 3 & 0 & 0 \\
2 & -1261.99 & 1088.49 & 16.410 $\pm$ 0.014 & 1.191 $\pm$ 0.026 & \nodata 
& \nodata & \nodata & 0 & 3 & 3 & 0 & 0 \\
3 & -1259.21 & -132.86 & 16.202 $\pm$ 0.008 & 0.768 $\pm$ 0.016 & \nodata 
& \nodata & \nodata & 0 & 3 & 3 & 0 & 0 \\
4 & -1259.00 & 995.52 & 16.799 $\pm$ 0.016 & 0.885 $\pm$ 0.020 & \nodata
& \nodata & \nodata & 0 & 3 & 3 & 0 & 0 \\
5 & -1256.87 & 648.49 & 12.586 $\pm$ 0.005 & 1.815 $\pm$ 0.009 & \nodata
& \nodata & \nodata & 0 & 6 & 4 & 0 & 0 \\

\enddata

\tablecomments{Table \ref{catalog} is published in its entirety in the 
electronic 
edition of the {\it Astronomical Journal}.  
A portion is shown here for guidance
regarding its form and content.}

\end{deluxetable}

\begin{figure}
\epsscale{1.05}
\plotone{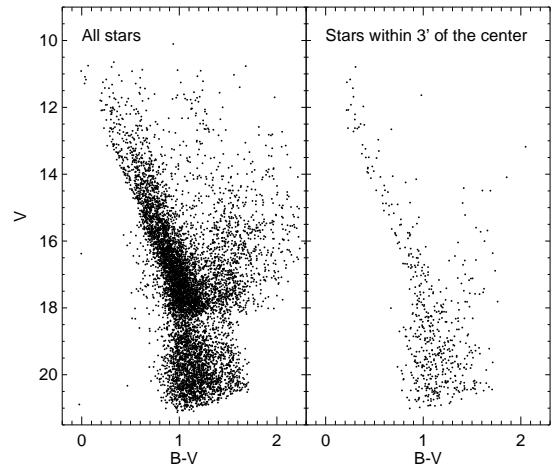}
\caption{(a) CMD for all stars in the catalog (field of view
about $43 \times 43$ 
arcminutes).  The apparent discontinuity at magnitude 18.5 arises because the 
Hall 1.2-m frames cover a wider field, but are not as deep.
(b) CMD 
for stars within a radius of 3 arcminutes from the cluster center only.
\label{allbv}}
\end{figure}

Figure \ref{allbv}a shows photometry for all
stars in the catalog, covering a $42.8^{\prime} \times 42.8^{\prime}$\ 
arcminute field, and Figure \ref{allbv}b is a \cmd\  derived from 
photometry of stars within a radius of 3 arcminutes from the cluster center.

\section{Cluster Structural Parameters}

\begin{figure}
\plotone{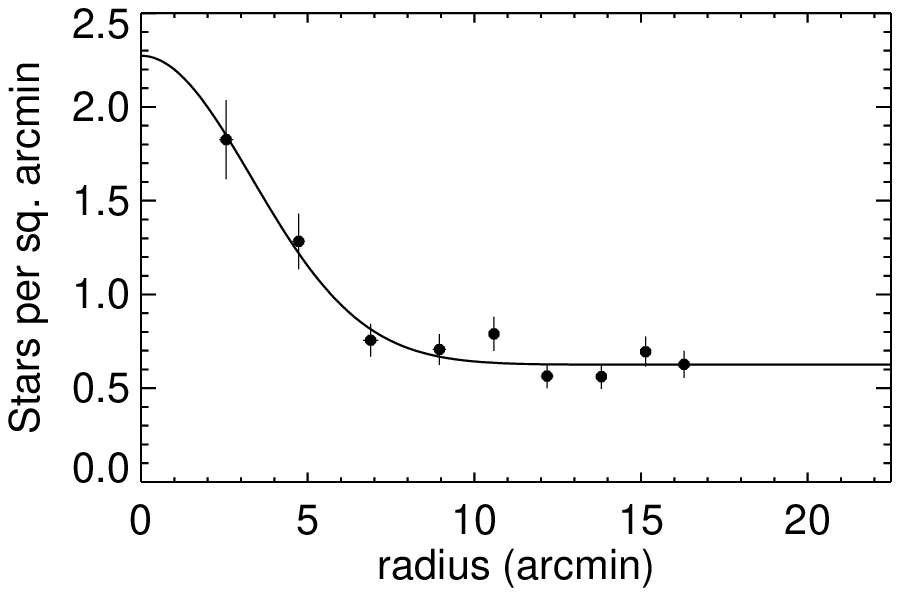}
\caption{Radial density distribution of stars in NGC 6866.  Each point 
represents the mean density of a ring containing 75 stars.  The curve is an 
exponential with $\sigma_r = 4.7$\ arcminutes.
\label{profile}}
\end{figure}

Using the data in Table \ref{catalog}, we derived the radial stellar 
density distribution about the cluster center.
We calculated the areas of successive rings 
within which there are 75 stars between magnitudes 11 and 16.  From that we 
found the stellar density in each ring, as shown in Figure \ref{profile}.
Because the cluster is relatively sparse and embedded in a rich star 
field, we chose a simple exponential 
function to model the radial stellar density
profile:
\begin{equation}
\rho(r) = \rho_{cl} \thinspace exp\left({-r^2 / \sigma_r^2}\right) + \rho_f,
\label{expfn}
\end{equation}
where, $\rho_{cl}$
refers to the peak of the radial 
stellar density distribution,
$\rho_f$ is the background field stellar density.  
and the cluster core angular radius is $\sigma_r$.

Assuming the stellar density profile of Equation \ref{expfn}, 
the effective angular radius is $\sigma_r$ = 4.7 arcminutes 
and the peak cluster
density is $\rho_{cl}$ = 1.65 stars per square
arcminute on a field background of 0.63 stars per square arcminute. 
After subtracting the background, the integral of 
the exponential density  function 
over all radii is
\begin{equation}
N_{cl} = \pi \rho_{cl} \thinspace \sigma_r^2,
\label{integral}
\end{equation}
which gives a cluster population of 114 stars brighter than magnitude 16. 
This should be considered to be a lower limit to the cluster membership
since there is likely to be an extended halo of stars photometrically
undetectable against the background field population.  Proper motions or radial 
velocities would be required to distinguish halo cluster stars from field stars;
fortunately the goals of this project require only that a sufficiently large
sample of high probability cluster stars be identified.

\section{A Bayesian Analysis of NGC 6866}

Bayes' theorem states that
\begin{equation}
P({\bf m}|{\bf D}) = {P({\bf m}) \mathcal{L}({\bf D}|{\bf m}) \over 
P({\bf D})},
\label{bayes}
\end{equation}
where $P({\bf m}|{\bf D})$\ 
is the posterior probability distribution function (PDF) of the 
model parameters, ${\bf m}$, given the data, ${\bf D}$. $P({\bf m})$\ 
is the prior PDF of the model 
parameters and $\mathcal{L}({\bf D}|{\bf m})$\ 
is a likelihood function for the data, given a model.  
The denominator, $P({\bf D})$\ 
is the integral of the numerator over all possible model 
parameters, and is ordinarily difficult to compute.  However, it is 
effectively a normalizing constant, and the shape of the posterior PDF can be 
found without knowing $P({\bf D})$.  
A thorough discussion of Bayesian analysis can be found in \citet{gregory05}; 
for some astronomical applications, including deriving properties of star 
clusters,
see \citet{ford05}, \citet{vonhippel06} or \citet{degennaro09}.  Our 
procedure is discussed in greater detail in Paper I.  The 
following is a brief summary of our cluster analysis procedure. 
 
{\it Priors} -- 
The prior distribution, $P({\bf m})$, 
constitutes a model for the cluster.  
The model parameters include global properties of distance modulus, reddening 
and a theoretical CMD of some age and metallicity, as well as the  membership 
probability and binary star status of individual stars. 
Because the global properties (distance modulus, reddening, age and
metallicity) are chosen with equal probability within some 
range, their contribution to $P({\bf m})$ is effectively a 
constant and does not affect
the shape of the posterior PDF. At each iteration, the model 
consisted of a theoretical CMD created from an 
isochrone of the specified age and metallicity by selecting a set of stars 
with random masses from the isochrone.  The theoretical CMD was corrected for
the particular values of reddening and distance selected at this iteration.
We ran two sets of models, one using the ``Yale-Yonsei'' (YY) isochrones 
\citep{demarque04} and the other using ``Padova'' isochrones \citep{bressan12}.
The models were allowed to range over values of $0.000 \le E(B-V) \le 0.30$;
$10.0 \le (m-M)_V \le 11.75; 300\ Myr \le$\ age $\le 1200\ Myr$; and 
$0.001 \le Z \le 0.030$.  All distance moduli calculated in this paper are 
observed, V-band moduli, uncorrected for interstellar extinction.

{\it Cluster Membership} -- We computed the prior membership probability for 
individual stars, based on their distance
relative to the cluster center using Equation \ref{expfn}. So the
prior membership probability for star $i$\ is a purely astrometric quantity, 
\begin{equation}
a_i \propto exp(-r_i^2 / \sigma_r^2).
\label{astrometry}
\end{equation}
At each iteration of the Markov chain, we randomly 
selected 125 stars from the entire field
with prior membership probabilities given by Equation \ref{astrometry}. That is
to say, after randomly selecting a star from the catalog, we compared its 
prior membership probability as derived 
from Equation \ref{expfn} with a randomly chosen number from zero to one.  If
the star's membership probability exceeds that of the random number, it is 
added to the sample for that iteration.  In this way, each star can be tested
as a possible cluster member; stars which are more often 
present in successful trials 
have a higher posterior membership probability, as discussed in the following.

{\it Binary Stars} --
The large fraction of binaries among open cluster stars makes 
it difficult to define the location of the actual single-star main sequence, 
particularly near the turnoff.  However, the binary status of individual stars 
can be added as parameters in the Bayesian analysis.  
Each star is assigned a binary mass ratio, and the merged magnitude and
colors of an equivalent  
binary in the theoretical CMD is calculated and compared to the observed star.
For the initial iteration, the binary mass ratio for all stars was taken to be
zero.  At each subsequent 
iteration, the mass ratio is randomly perturbed from the previous value
with a Gaussian probability (see the Hastings-Metropolis algorithm below).

{\it Likelihood Function} -- 
For each observed star, $i$, we find the distance, $z_{i,j}$\ in 
units of the measurement errors in the 5-dimensional CMD space (V, B-V, U-B, 
V-R, V-I) to each of the theoretical stars, $j$.  The probability that 
$z_{i,j}$ is 
as small as it is by chance is the error function of $z_{i,j}$, 
$erf(z_{i,j})$. The 
probability that the observed star could be drawn from the CMD of theoretical 
stars is given by
\begin{equation}
p_i = 1.0 - \prod_j\Bigl[erf(z_{ij})\Bigr].
\label{starprob}
\end{equation}
So now from Equations \ref{bayes} and \ref{starprob}, 
the operational Bayesian model is 
\begin{equation}
P({\bf m}|{\bf D}) = P({\bf m}) \mathcal{L}({\bf D}|{\bf m}) =
P({\bf m})\prod_i p_i,
\label{posterior}
\end{equation}
which gives the probability that the observed sample of stars could be drawn
from the theoretical sample.

{\it Hastings-Metropolis Algorithm} –- 
The key to efficient sampling of the PDF is to 
develop a “jump probability function,” $\alpha ({\bf m_t}|{\bf m})$, 
to decide the transition from 
model ${\bf m}$ to a new trial 
model ${\bf m_t}$ so that ${\bf m_t}$ depends on the current 
parameters but not on any
previous values. The trial model, ${\bf m_t}$, 
consists of random proposed jumps, $q({\bf m})$, from the 
current parameter values. 
The jump probability 
function can be written as 
\begin{equation}
\alpha({\bf m}_t | {\bf m}) = min\bigl\{{P({\bf m}_t | {\bf D})\thinspace
\over P({\bf m} | {\bf D})} {q({\bf m} | {\bf m}_t) \over
 q({\bf m}_t | {\bf m})} , 1\bigr\}.
\label{markov}
\end{equation}

If the jumps have a Gaussian distribution,
the proposal distributions are symmetrical and in this case,
$q({\bf m} | {\bf m}_t) = q({\bf m}_t | {\bf m})$.  
A trial model is accepted as the next iteration when $\alpha$\ 
exceeds a random number 
in the range 0 to 1; if the trial model is not accepted, another
${\bf m}_t$\ is tested.  
Improved solutions are always selected, but in addition, 
the entire parameter space is eventually sampled. 
We ran two chains of 105000 trials; after throwing away the first 5000 trials
in the ``burn-in'' period, we saved all trials satisfying the 
acceptance condition on the 
first try.  See Paper I for more information.

\begin{figure}
\epsscale{1.1}
\plotone{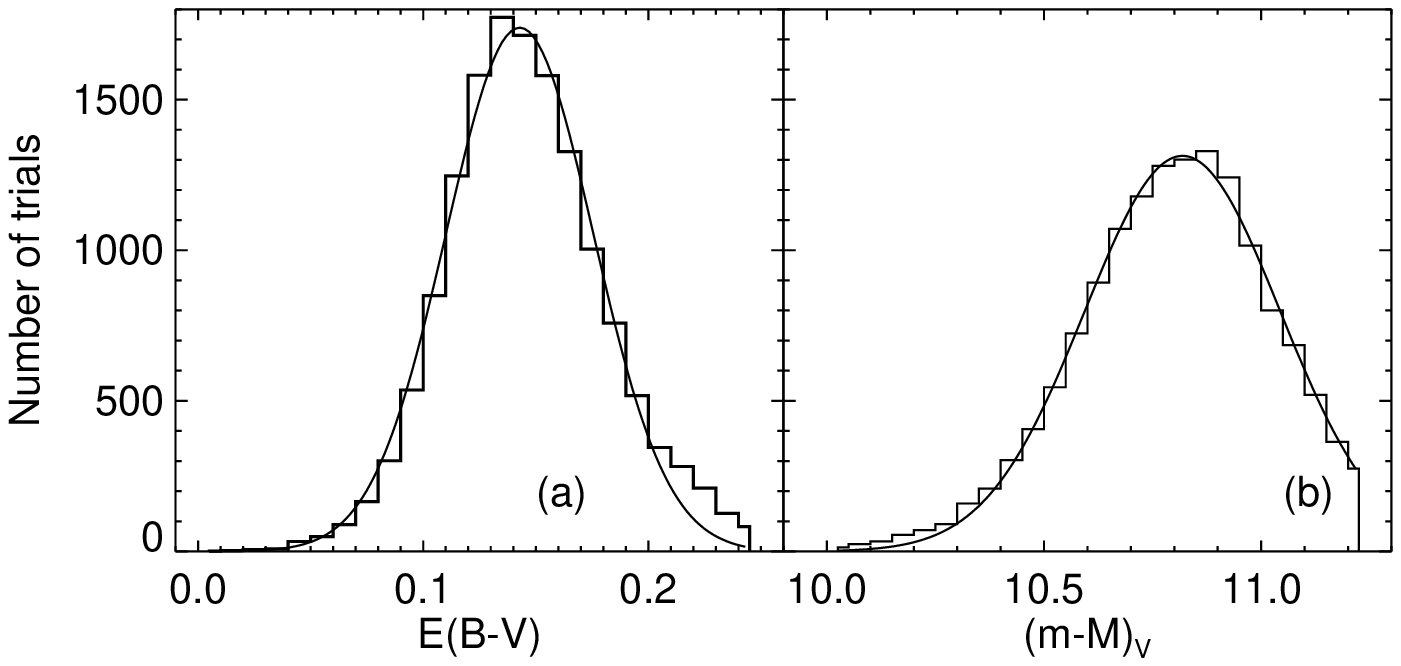}
\epsscale{1.1}
\plotone{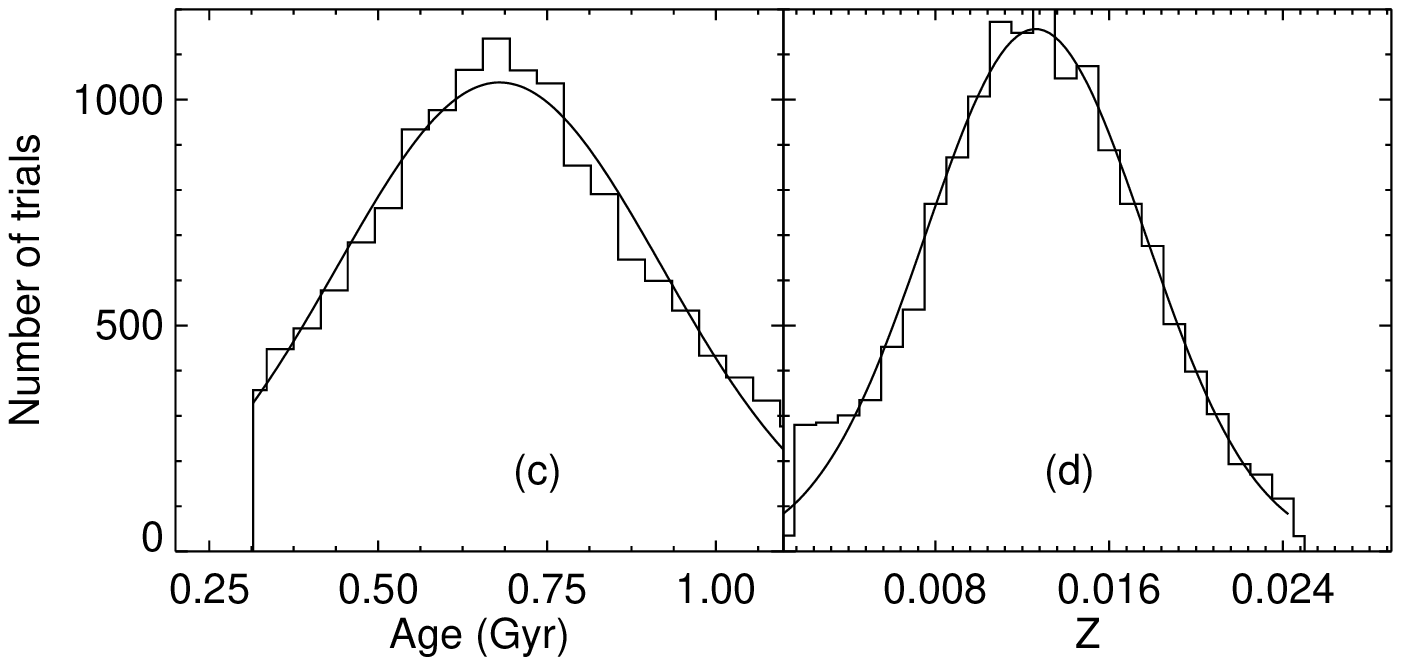}
\caption{Marginal distributions of cluster properties for the Yale chain:
Number of samples
vs: (a) reddening, E(B-V), (b) distance modulus, (c) age in Gyr and (d) 
metallicity, Z. 
In each figure, the 
solid curve represents a Gaussian fit to the distribution.
\label{margins}}
\end{figure}

\section{Discussion}

The product of the above analysis is the posterior PDF, $P({\bf m}|{\bf D})$, 
a sequence of the 
parameter values for each successful trial.  
From the marginal distribution
of each parameter (found by integrating $P({\bf m}|{\bf D})$
over all the other parameters), 
its mean and variance can be found.  The marginal distributions are shown
in Figure \ref{margins}; although the scale of the distributions are
defined only by the unknown parameter $P({\bf D})$, their means
represent the maximum likelihood values of the parameters, and their 
standard deviations represent the parameter uncertainties.
These are shown in 
Table \ref{final}.  

\begin{figure}
\epsscale{1.1}
\plotone{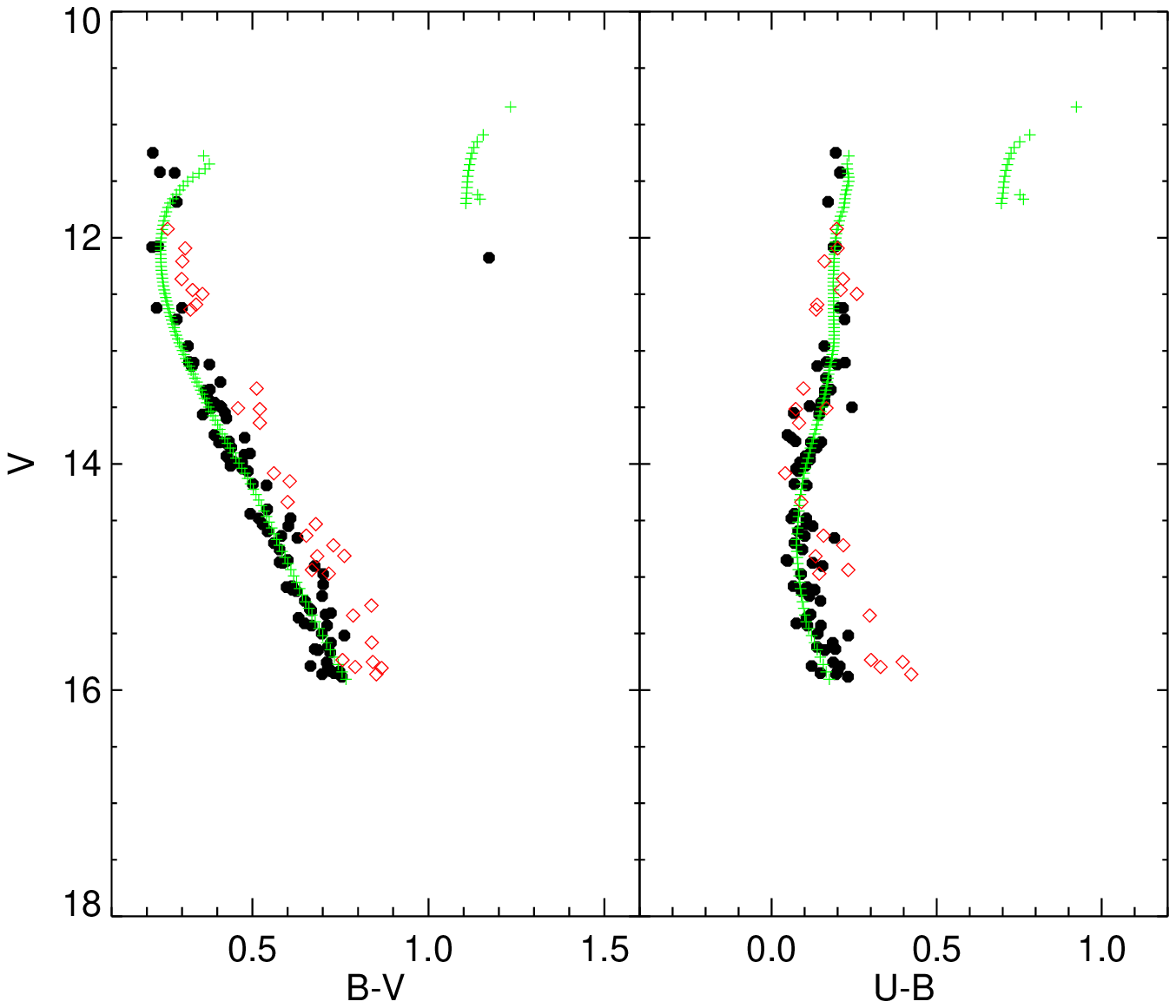}
\epsscale{1.1}
\plotone{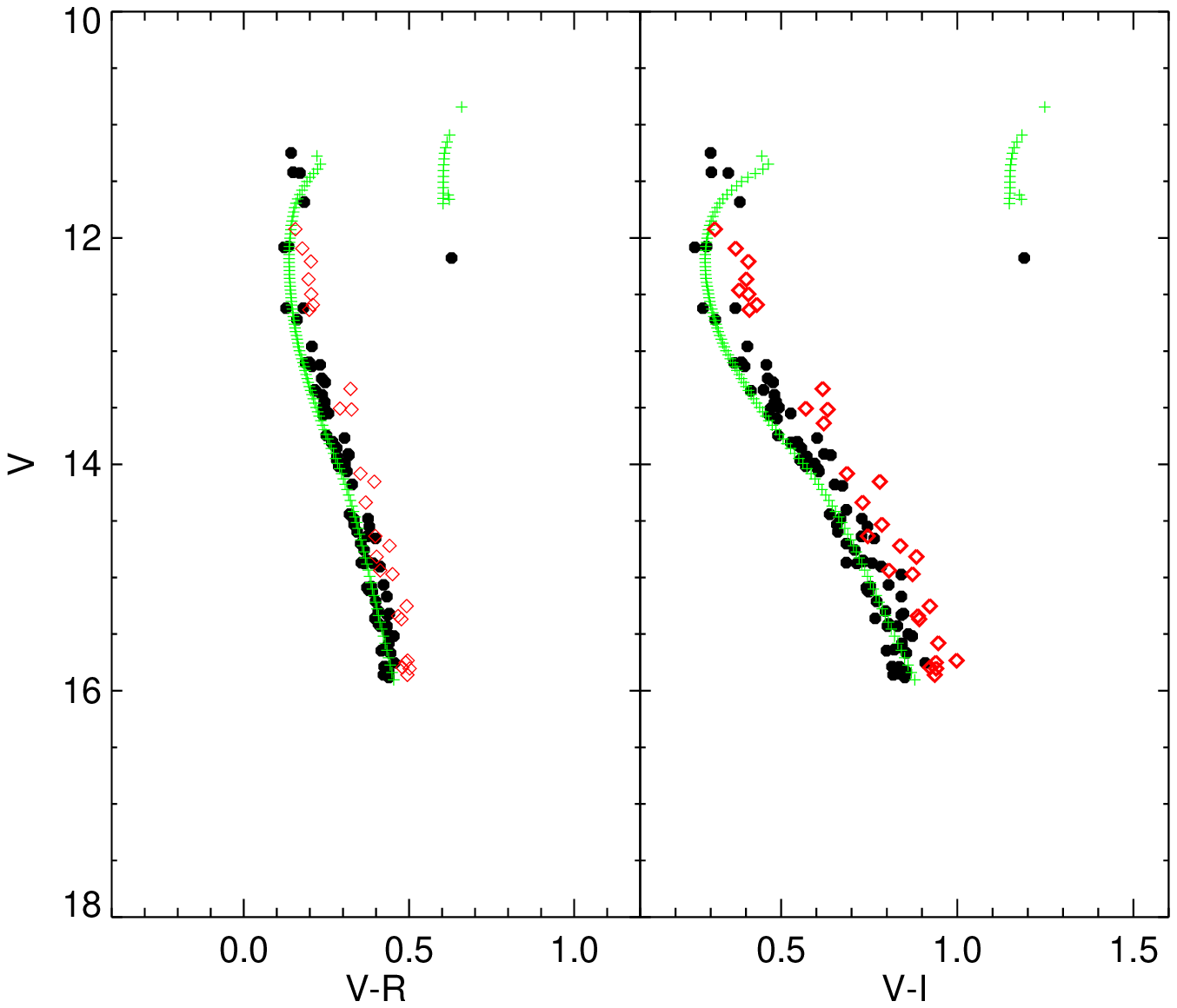}
\caption{CMDs of the 125 stars in the cluster region
with the largest posterior membership 
probabilities. Filled symbols: likely single stars; Open symbols: probable 
binary stars. Small crosses:
theoretical CMD derived from the maximum-likelihood Padova isochrone. 
See the online edition 
for a color version of this figure.
\label{cmdfits}}
\end{figure}

Up to a constant, the prior PDF of an individual star is a single number, 
$a_i$, given by Equation \ref{astrometry}.  So now, the posterior PDF of 
star $i$\ is just the product of $a_i$\ and the individual likelihood
parameter, $p_i$, given by Equation 
\ref{starprob}.
In Figure \ref{cmdfits} we included 125 stars with the largest values of the
posterior membership probability.

For intermediate-age clusters like NGC 6866, with few if any giant stars, there
is little information in the CMD to distinguish clusters by age or metallicity.
This problem is intrinsic to isochrone fitting to 
clusters in this age range, independent of 
the details of the analysis technique.  
Furthermore, NGC 6866 is embedded in a rich field.  As a consequence, 
although the procedure produces well-defined posterior PDF's, the 
marginal distributions are rather broad –- i.e, the parameters are only 
moderately well constrained. 

The consequences of the
indeterminancy between the effects of reddening and metallicity on
photometry can be seen in Figure \ref{twod}, a contour plot of the 
marginal distribution of the number of samples in the E(B-V) vs Z plane.
Solutions with low reddening values tend to be correlated with 
those with  high
metallicity values.  
If one or the other were constrained by independent 
measures, the marginal distributions of the other parameters would be
considerably sharper. There are indications of similar correlations between
the other indices, but they are not as strong.

\section{Conclusions}

We have produced a catalog of high-quality Johnson/Cousins UBVRI photometry
of stars in a large field around the cluster NGC 6866.  Our photometry has
been carefully transformed to the standard system as defined by the 
\citet{landolt09} standard stars.  Using a Bayesian analytical procedure, 
we derived the following values for the cluster properties:
$(m-M)_V = 10.98\pm 0.24$, $E(B-V) = 0.16\pm 0.04$,
age $= 705\pm170$\ Myr and
$Z = 0.014\pm 0.005$ (see also Table
\ref{final}). Assuming a canonical
extinction-to-reddening ratio $R_V = A_V / E(B-V) = 3.1$, 
the distance to 
the cluster is about 1250 parsecs.

\begin{figure}
\plotone{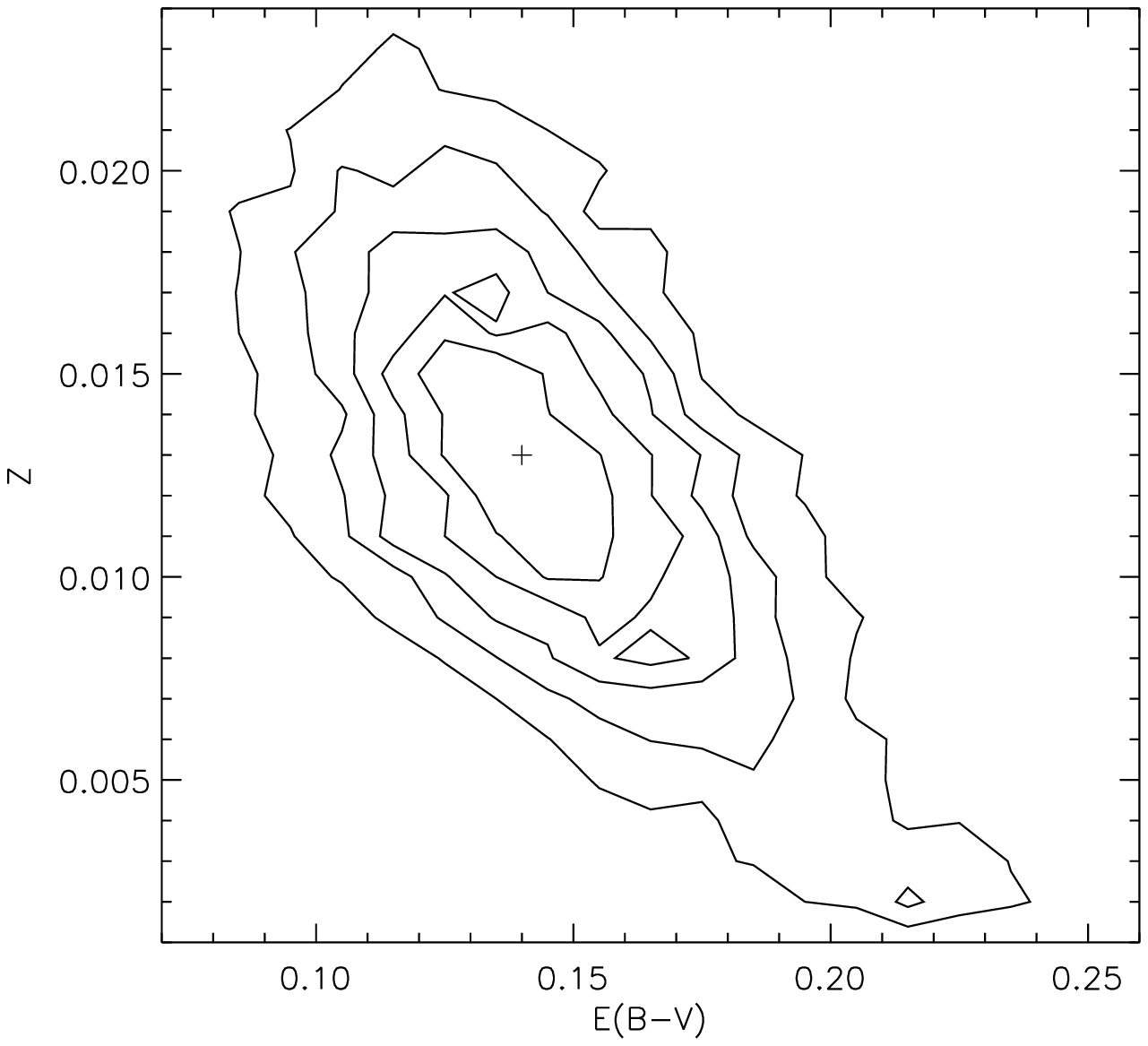}
\caption{Two-dimensional marginal distribution of the numbers of trials in
the E(B-V), Z plane for the Yale chain.  The plus sign marks the peak of the
Yale chain distributions for E(B-V) and Z (see Table \ref{final}).
\label{twod}}
\end{figure}

Our results are in rough agreement with previous work on this cluster as
discussed in the introduction 
\citep[see also][and references therein]{balona13} 
As mentioned above, the lack of independent membership 
information, or, alternatively the cluster metallicity or reddening, 
limits the accuracy of photometric measurements of its 
age and distance. Furthermore, most of the previous results were based, 
directly or 
indirectly, on visual comparisons of observed color-magnitude diagrams with 
theoretical isochrones of solar composition only; for that reason, 
the quoted errors are likely to be highly optimistic, since they do not include
the metallicity uncertainty.  In particular, our analysis, as shown in 
Figures \ref{margins} and \ref{twod}, do not support a metallicity as high
as solar.  Solar metallicity would require a reddening of essentially zero,
unlikely in this direction in the galaxy.  

Finally, the errors given in Table 3 do not include any uncertainty in
the theoretical models.  In particular, Figure \ref{cmdfits} suggests that 
there may be a small 
inconsistency between the theoretical models and the observed CMD.

\begin{deluxetable}{ccccc}
\tabletypesize{\small}
\tablewidth{0pt}
\tablecaption{NGC 6866 --- Summary of MCMC Analysis
\label{final}}
\tablehead{\colhead{ } &\colhead{E(B-V)} & \colhead{$(m-M)_V$}   
& \colhead{Age (Myr)} & \colhead{Z}}

\startdata

Padova &0.17$\pm$0.05 & 11.15$\pm$0.36 & 730$\pm$320 & 
0.014$\pm$0.008 \\
Yale-Yonsei &0.14$\pm$0.05 & 10.82$\pm$0.32 & 680$\pm$340 &
0.013$\pm$0.007 \\
Average &0.16$\pm$0.04 & 10.98$\pm$0.24 & 705$\pm$170 & 0.014$\pm$0.005 \\

\enddata

\end{deluxetable}

\acknowledgments

We 
want to acknowledge the financial and technical support from 
Boston University and Lowell Observatory. SAB is grateful for financial 
support from the Barnes and LoMonaco families, and thanks the Flagstaff Public
Library for providing a serene working environment during a crucial phase of
this work.
SM acknowledges support from NASA cooperative 
agreement NNX09AH18A (the Kepler Cluster Study).

Facilities: \facility{Perkins (PRISM)}

\bibliographystyle{apj}

\clearpage


\end{document}